\documentstyle[11pt,newpasp,twoside,epsf]{article}
\def\arg#1{{\rm#1\/}}

\def\edcomment#1{\iffalse\marginpar{\raggedright\sl#1\/}\else\relax\fi}
\reversemarginpar

\begin{document}
\title{Gas flow in barred galaxies }
\author{E. Athanassoula }
\affil{Observatoire de Marseille, 2 Place Le Verrier,
13248 Marseille cedex 04, France}

\begin{abstract}
I briefly review the properties of the gas flow in and around the region
of the bar in a disc galaxy and discuss the corresponding inflow and
the loci of star formation. I then review the flow of gas in
barred galaxies which have an additional secondary bar. Finally I discuss the 
signatures of bars in edge-on galaxies.
\end{abstract}

\section{Introduction}

Bars are elongated structures frequently present in the
central parts of disc galaxies. Their formation is ubiquitous 
in N-body simulations, unless a sufficiently massive and sufficiently
centrally concentrated spherical or spheroidal component has been
added to stop or at least to delay their formation beyond a reasonable
life-time of the disc. The problems relating to bar formation are far from
being solved, but in this review I will leave them aside in order to
concentrate on the properties of the gas flow in and around the bar region. 
Such a flow is intimately linked to the main periodic orbits and their
structure (e.g. Athanassoula 1992a; Athanassoula 1992b, hereafter
A92b). Orbital calculations 
(for a review see e.g. Sellwood \& Wilkinson 1993) have
shown that the backbones of bars are the so-called $x_1$ periodic
orbits, which are elongated along the bar. If one or two inner Lindblad
resonances (hereafter ILRs) are present, there are also two
perpendicular families, called $x_2$ and $x_3$. Further out we find the
4:1 periodic orbits, which have rectangular-like or diamond-like
shapes. These families, as well as other secondary ones, 
trap around them a number of regular orbits. 
Chaotic orbits are also present and their importance depends on the
properties of the bar potential.

\section{Gas flow in and around the bar }

\begin{figure}
\plotfiddle{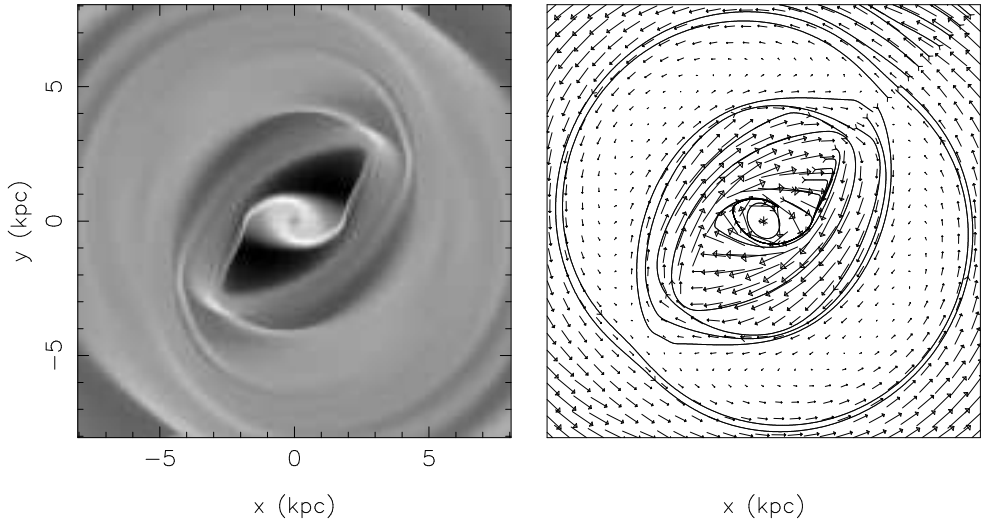}{7cm}{-90}{90}{90}{-350}{300}
\caption{Response of the gas to a bar model. The left panel shows the
gas density in grey-scale and the right panel the gas flow lines and
the velocity vectors in a frame of reference in which the bar is at
rest. The side of each box is 16 kpc.}
\label{fig:fig1}
\end{figure}

An example of the gas response to a barred galaxy potential, calculated
with the FS2 code (van Albada, van Leer \& Roberts 1982; van Albada,
1985), is given in 
Fig.~1. The left panel gives the density response in
grey-scale; lighter shades correspond 
to higher densities and darker to lower ones. The imposed bar is at
45\deg~ 
to the horizontal axis, rotates clock-wise and has a semi-major axis
of 5 kpc and a 
semi-minor one of roughly 2.3 kpc. We note that in most of the bar region the
gas has
very low density, while most of the gas is concentrated in two narrow
strips along the leading edges of the bar. In their inner parts these
density enhancements curve into a nuclear spiral or ring-like
structure. There is also considerable
gas concentrated around the extremities of the bar major axis. The
right panel shows the gas 
flow in the same model and in a frame of reference corotating with the
bar. The length of 
the vectors is proportional to the velocity at that location and I have
also superposed a number of flow lines. Comparison of the two panels,
as well as cuts perpendicular to the density maxima (cf. A92b),
show that the density maxima are the loci of shocks. Prendergast (1962,
unpublished) was the first to associate them with the dust
lanes observed along the leading edges of bars. A comprehensive study
of the shape of the shock loci and of their dependence on the main
parameters of the model can be found in A92b. The shock loci in strong
bars were shown to be straight, while weak bars or ovals have loci
which are curved
with their concave part towards the bar major axis. One of the main results of
A92b is that, in order for the shock loci to have the shape of
the observed dust lanes, the corotation radius, $R_{CR}$, must be
equal to $R_{CR}=(1.2\pm0.2)a$, where $a$ is the length of the bar
semi-major axis. 

A further result of A92b is that, in order for shocks to exist, 
the curvature of the $x_1$ orbits at apocenter must exceed a certain
quantity, or in other words, the orbits that are elongated along the
bar should be sufficiently peaked or have loops at their
apocenters. Furthermore, in 
order for the shock loci to be offset towards the leading side of the
bar, the $x_2$ orbits must not only exist, but also have a sufficient
extent. 

\subsection{Inflow}

In cases with no shocks the flow lines have simple concentric
ellipse-like shapes, and there is no net inflow.
However, in cases with shocks the gas flow is considerably more
complicated, as can be seen in the right panel of Fig.~1. At the
trailing sides of the bar, where the gas density is very low, the flow
is outwards, and stays so until it reaches the shock. At that point it
turns abruptly inwards. Both the outwards and the inwards flow reach
high values of the velocity, of the order of, or higher than, 100
km/sec. The outwards 
flow is over an area which is large, but has a very low gas density, while the
inwards flow is concentrated around the density maxima. i.e. in a
small area with very high gas density. Thus the net inflow,
i.e. the density averaged radial velocity, is of the order of less
than a km/sec, up to a few km/sec, depending on the model.
Care has to be taken when comparing with observations since
the low density regions are harder to observe and thus there is a
tendency for observations to overestimate the inflow. 

A maximum net inflow is found in models with thin, massive, slowly
rotating bars, with centrally concentrated axisymmetric
components (A92b; Piner, Stone \& Teuben 1995, etc). Since these models have
also the strongest shocks (A92b),
we come to the conclusion that the most important inflow is in models with
strongest shocks, as expected.

\subsection{Can the inflowing gas reach the nucleus? }
\label{sec:in}

In cases with ILRs the gas is brought from regions roughly within a radius
equal to the bar semi-major axis to the circumnuclear ring or
pseudo-ring 
and little to the disc within it. On the other hand in cases with no ILR 
the inflow reaches the center-most area, i.e. very near the
nucleus. There are indications, however, that most
barred galaxies have one or two ILRs (Athanassoula 1994),
so that in general the bar will not push the gas sufficiently
inwards to reach the nucleus and eventually feed an AGN.

Nevertheless even in cases with ILRs gas can be
pushed to the center-most radii. One possibility, which will be discussed in
section~3, is that there is a 
second bar within the main one. Other mechanisms for breaking the ILR
barrier rely on the 
self-gravity of the gas accumulating in the central area. Such
mechanisms have been described e.g. by Fukunaga \& Tosa
(1991), Wada \& Habe (1992, 1995), Elmegreen (1994) and Heller \&
Shlosman (1994). 

\subsection{Star formation }

The shock loci are regions of high density maxima, and one could have
naively thought that this might entail a considerable amount of star
formation. This, however, is not the case, due to the fact
that straight shock loci are
also the loci of high shear (A92b). Thus a typical molecular cloud
will shear out before it has time to collapse, so that no star
formation will occur. This is not necessarily true for shocks whose loci
are curved. The two different types can be seen when comparing e.g. the
dust lanes in the bar of NGC~1300 with the corresponding ones in
NGC~1566. With a similar goal,
Regan, Vogel \& Teuben (1997) stressed the large divergence in gas
streamlines before they reach the shock, and argued that this could
tear apart molecular clouds and thus inhibit star formation. It would
be interesting to test whether
this mechanism works for straight dust lanes as opposed to
curved ones, as observations seem to suggest. 

The ends of the bar are regions of high density, but 
not much shear. Thus one expects to find star formation there and
this is indeed borne out by observations. 
Strong gas concentration can also be found in the area covered by the
$x_2$ orbits, if the 
galaxy is sufficiently centrally concentrated (e.g. early type disc
galaxies) and if the bar is sufficiently strong (thin and/or
sufficiently massive). In such cases star formation can occur in the
nuclear pseudo-ring, or in the nuclear disc.

\begin{figure}
\plotfiddle{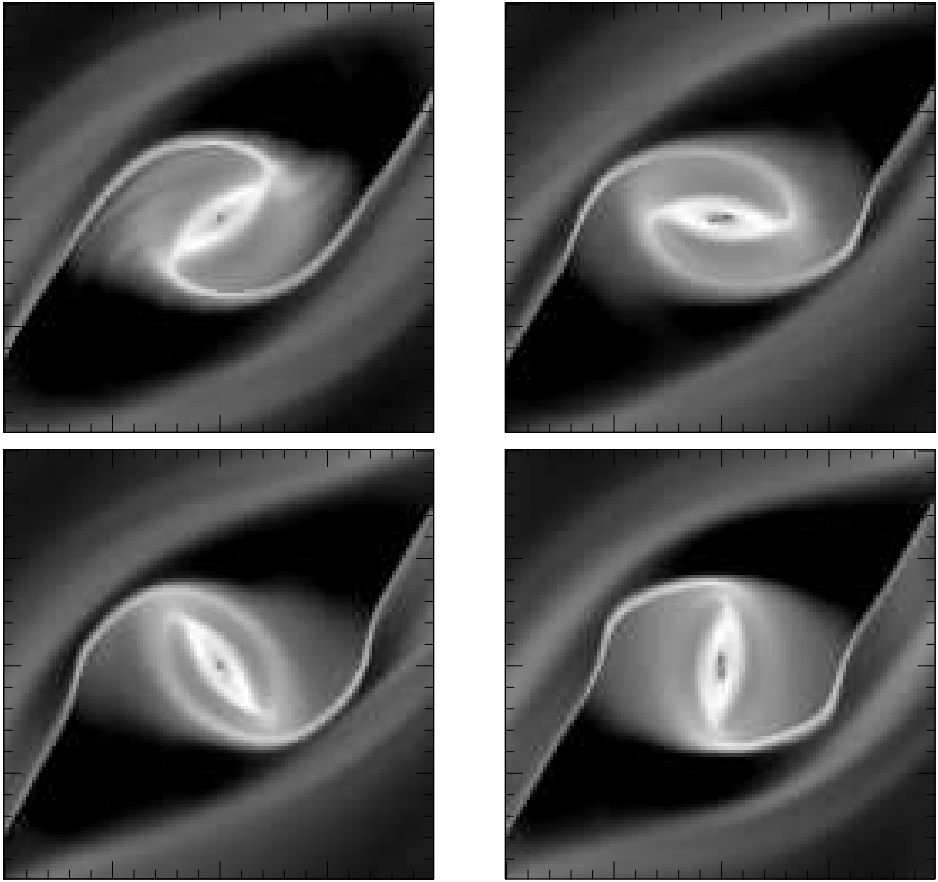}{12.0cm}{-90}{100}{100}{-190}{370}
\caption{Response of the gas in the inner 2 kpc $\times$ 2 kpc area of
the bar, at four different times during the simulation described in
section 3. As in Fig. 1, lighter areas correspond to high density regions
and dark ones to low density ones. }
\label{fig:fig2}
\end{figure}

\section{Bars within bars }
\label{sec:binb}

A large number of observations have shown the existence of inner bars
in different components, as the CO, the visual, or the NIR.
Their formation could be due either to the existence of a bar
unstable stellar inner disc (i.e. an instability similar to that forming the
main bar), or to the fact that sufficient gas has been pushed inwards
to form a bar unstable gaseous disc (Shlosman, Frank \& Begelman
1989). The main bar and the inner 
bar do not necessarily turn with the same pattern speed, but this does
not mean that they are dynamically independent. Indeed the two pattern
speeds can be different but 
coupled. Analytical work has predicted that the location of the ILR of
the main bar 
should coincide roughly with that of the corotation of the inner bar (Tagger
et al. 1987, Sygnet et al. 1988), and this
has been nicely borne out by the simulations of Friedli \& Martinet
(1993). As already mentioned in section~2.2 such inner bars can
help push the gas to the innermost regions of the galaxy. The mere
existence of an inner bar may nevertheless not be sufficient for 
this task. 

The response of the gas in cases with two imposed bars will of
course vary with the angle between the major axes of the two
bars. Thus when the two bars have different pattern speeds the response
is a function of time. This is illustrated in Fig.~2, where I display
the nuclear parts of the gas response at four different times during a
simulation. The outer (or main) bar is always at 45\deg~ 
to the horizontal axis and has a semi-major axis of 5~kpc. The
axisymmetric component has been chosen such that the rotation curve
rises very steeply in the inner parts. The gas response in the model
with only the outer bar 
has strong shocks along the leading sides of the bar that, towards
their innermost parts, wind up in the form of a nuclear spiral, as is
the case for the model shown in Fig.~1. It is
this nuclear region that gets most affected by the inner bar, as
expected. In the example shown in Fig.~2 the inner
(or secondary) bar rotates roughly
3.5 times faster than the main bar, so that the corotation of the
inner bar is situated roughly at the ILR of the main one. The gas
response in the innermost parts is now bar shaped, but a given end of
this gaseous bar does not always link to the same density enhancement
of the outer bar. The gas response adjusts itself so that a given side
of the inner
gaseous bar links to the nearest density enhancement within the main bar,
and in some cases one sees a nuclear pseudo-ring surrounding the inner
gaseous bar.  
The work described briefly here is part of an extensive study of the
response of gas in galaxies with both a primary and a secondary bar,
which I have been doing in collaboration with G. D. van Albada. A more
comprehensive description of this work will be published elsewhere.

\section{Bars in edge-on galaxies }
\label{sec:edgeon}

N-body simulations show clearly that stellar bars do not stay flat thin
structures, but extend vertically, taking the form of a peanut if seen
along the bar minor axis, and a box-like shape if seen edge-on and at
an angle to 
it (Combes \& Sanders 1981, Combes et al. 1990, Raha et al. 1991 etc). 
Such peanut-like or box-like protuberances are observed in many disc
galaxies and are called peanut or boxy bulges. It is
not easy to prove that they are indeed bars seen edge-on by using photometry
alone. Thus efforts 
have been concentrated on kinematics.  Kuijken \& Merrifield (1995)
and Bureau \& Freeman (1999) have compared the position velocity
diagrams (PVDs) of galaxies with peanut or box-like bulges, to the PVDs
of galaxies with standard bulges and find importance differences,
which, they argue, could well be due to
the signatures of the $x_1$ and $x_2$ families of
orbits in the peanuts. Bureau \& Athanassoula (1999) 
made a detailed study of the effect of the existence and
properties of the
families of periodic orbits on the structure of the PVDs.
Athanassoula \& Bureau (1999) have tackled the same problem
using hydrodynamical simulations, and they find that the existence of a
gap in a PVD, between the signature of the nuclear spiral and that
of the outer disc, reliably indicates the presence of a bar. This gap
is due to the fact that most of the main bar region is relatively
empty, and this in turn is due to the gas flow described in section
2. All arguments seem to converge to the 
fact that peanut and boxy bulges are misnamed, and are in fact bars
seen edge-on.

\end{document}